\documentclass[prl,aps,superscriptaddress,nofootinbib,amsmath,amssymb,nopacs,twocolumn]{revtex4-2}
\usepackage{graphicx}
\usepackage{braket}
\usepackage{nicefrac}
\usepackage{float}%
\usepackage{amsmath}
\usepackage{hyperref}
\usepackage{multirow}
 \usepackage{bbold} 

\usepackage[usenames, dvipsnames]{color}

\def\slashchar#1{\setbox0=\hbox{$#1$}
   \dimen0=\wd0 \setbox1=\hbox{/} \dimen1=\wd1
   \ifdim\dimen0>\dimen1 \rlap{\hbox to \dimen0{\hfil/\hfil}} #1
   \else  \rlap{\hbox to \dimen1{\hfil$#1$\hfil}} / \fi}

\graphicspath{{./plots/}}

\begin{document}
\title{Nuclear medium and final state interaction effects in $\eta$ production induced by $\nu_\mu(\bar\nu_\mu)$ at SBN and DUNE energies}

\author{A. \surname{Fatima}}
\affiliation{Department of Physics, Aligarh Muslim University, Aligarh-202 002, India}\author{S. K. \surname{Noor Alam}}
\affiliation{Department of Physics, Aligarh Muslim University, Aligarh-202 002, India}
\author{M. Sajjad \surname{Athar}}
\email{sajathar@gmail.com}
\affiliation{Department of Physics, Aligarh Muslim University, Aligarh-202 002, India}
\author{S. K. \surname{Singh}}
\affiliation{Department of Physics, Aligarh Muslim University, Aligarh-202 002, India}

\begin{abstract} 
We investigate nuclear-medium modification and final-state interaction in charged- and neutral-current $\nu_\mu(\bar{\nu}_\mu)$-induced $\eta$-meson production on $^{12}\mathrm{C}$ and $^{40}\mathrm{Ar}$. The nuclear medium is described within a local density approximation, incorporating in-medium modifications of the $N^*(1535)$ resonance, which is known to play the dominant role in $\eta$ production through its $N \eta$ decay channel. The subsequent propagation and absorption of the $\eta$ meson in the nucleus are simulated using a Monte Carlo approach. We find that nuclear-medium modifications and final state interaction lead to substantial suppression of the production cross section, highlighting the importance of nuclear effects in interpreting $\eta$ production data. Our results provide valuable theoretical input for the analysis of $\eta$-production at the ongoing SBND and MicroBooNE experiments and the forthcoming DUNE experiment.
\end{abstract}
\maketitle

\textit{ Introduction - }
Neutrino interactions with nuclei constitute one of the dominant sources of systematic uncertainty in analyzing current and future accelerator based neutrino oscillation experiments which are being performed using moderate to heavy nuclear targets. A better understanding of these interactions over a broad energy range is essential for reliable neutrino energy reconstruction and for achieving the precision required in the determination of the neutrino oscillation parameters. In particular, the study of the exclusive meson production processes provide valuable information on the underlying weak hadron dynamics and on the modifications induced by the nuclear environment, thereby serving as sensitive probes of both neutrino interaction mechanisms and nuclear effects.

Among the various exclusive reaction channels, the weak production of $\eta$-mesons induced by the charged and neutral current occupies an important
position. Recent measurements of $\eta$ production by the MicroBooNE~\cite{MicroBooNE:2023ubu} and SBND~\cite{Linyan:2026, Alexander:2026} collaborations have further highlighted the need for quantitative theoretical predictions that consistently include both nuclear medium effects~(NME) and final state interaction~(FSI) of $\eta$ meson. Such studies are equally important for DUNE~\cite{DUNE:2020jqi} experiment, where exclusive meson production channels contribute to signal and background processes, as well as for atmospheric neutrino experiments like Super-K~\cite{Super-Kamiokande:2024qbv} employing nuclear targets. Owing to the strong coupling of the $\eta N$ channel to the odd parity spin-$\frac{1}{2}$ $N^*(1535)$ resonance, the weak $\eta$ production process provides a clean probe of resonance dynamics and offers direct sensitivity to the in medium properties of $N^*(1535)$ resonance. Inside nuclei, however, the strong $N\eta$ interaction is significantly modified by many body effects. The resonance propagates in a nuclear environment where its mass and width are altered by the surrounding medium. Moreover, the outgoing $\eta$ meson undergoes multiple elastic and inelastic interactions before emerging from the nucleus. These nuclear medium modifications and FSIs substantially affect the observable cross sections and must therefore be incorporated consistently in any realistic description of neutrino induced $\eta$ production.

To the best of our knowledge, this is the first calculation of charged-current (CC) and neutral-current (NC) $\nu_\mu(\bar{\nu}_\mu)$-induced $\eta$ production on nuclear targets. We consider $^{12}$C and $^{40}$Ar nuclei, which are of direct relevance to the scintillator and liquid-argon time projection chamber (LArTPC) detectors employed by the short baseline neutrino~(SBN) program, which comprises of three experiments viz. SBND~\cite{Linyan:2026, Alexander:2026}, MicroBooNE~\cite{MicroBooNE:2023ubu} and ICARUS~\cite{ICARUS:2023gpo}, and also for the planned DUNE~\cite{DUNE:2020jqi} experiment. For $\nu_\mu(\bar{\nu}_\mu)$ interactions on free nucleons, we employ the framework developed in Refs.~\cite{Fatima:2023fez, Fatima:2022nfn, Fatima:2025tht, SajjadAthar:2022pjt}, which provides a unified description of the weak production of the pseudoscalar mesons. We extend this framework to nuclear targets by incorporating nuclear-medium effects within the local-density approximation through the in-medium modification of the $N^*(1535)$ resonance. The propagation of the produced $\eta$ meson is treated using a Monte Carlo simulation that accounts for its absorption and the  FSI effects in the nuclear medium.

\textit{ Formalism - }
The CC and NC (anti)neutrino induced single $\eta$ production off the nucleon
target are given by the following reactions
\begin{eqnarray}\label{Ch12_eq:eta_weak_process_cc}
\nu_\mu/\bar{\nu}_{\mu} (k)  + N (p) &\longrightarrow& \mu^-/\mu^+ (k^\prime) + \eta ( p_\eta) + N^\prime (p^\prime),   \\
\label{Ch12_eq:eta_weak_process_cc1}
\nu_\mu/ \bar \nu_\mu (k)  + N (p) &\longrightarrow& \nu_\mu/ \bar \nu_\mu (k^\prime) + \eta( p_\eta)  + N (p^\prime) ,
\end{eqnarray}
where $N,N^\prime = n, p$ and  the quantities in the parentheses are the four momenta of the particles.

In the laboratory frame, the double differential scattering cross section  is expressed as~\cite{Fatima:2022nfn}:
\begin{equation}\label{sigma:weak}
 \left(\frac{d^2\sigma}{dW dQ^2} \right)_{\nu N}= \int dE_{\eta} \frac{1}{(2\pi)^{4}} \frac{\pi W}{32 M^2E_{\nu}^2} \frac{1}{|\vec{q}\;|} \overline{\sum} \sum |{\cal M}|^2,
\end{equation}
 where $W=\sqrt{(p+q)^2}$ is the center of mass~(CM) energy, $Q^2(=-q^2) \ge 0$ is the four momentum transfer square with $q=k-k^\prime$. ${\cal M}$  is given by
\begin{equation}\label{eq:mat:weak}
 {\cal M} = \frac{G_{F} a}{\sqrt{2}}  l_{\mu} J^{\mu},
\end{equation}
where $a=\cos\theta_C~(1)$ corresponds to CC~(NC) processes, $G_{F}$ being the Fermi coupling constant and $\theta_{C}$ being the Cabibbo mixing angle.
The leptonic current $l_{\mu}$ is given 
\begin{equation}\label{lepton:current}
 l_{\mu} = \bar{u} (k^{\prime}) \gamma_{\mu} (1\mp\gamma_5) u(k)
\end{equation}
where $-(+)$ stands for neutrino~(antineutrino) induced reactions 
and $J^{\mu}= J^{\mu}_{NR} + J^{\mu}_{R}$ is the weak hadronic current, which receives contribution from both the nonresonant~(NR) Born terms as well as the resonance~($R$) excitations~\cite{Fatima:2022nfn, Fatima:2023fez}.

The contributions from the $\eta$ production due to NR Born diagrams are calculated using a microscopic model based on the SU(3) symmetric chiral Lagrangian assuming $\eta$ belonging to the octet representation of SU(3). The parameters of the model are phenomenologically determined from the data available for the semileptonic decays of hadrons, and the details are given in Ref.~\cite{Fatima:2022nfn, Fatima:2023fez}.
The contribution from the resonant diagrams due to the excitation of various resonances like $N^\star(1535)$, $N^\star(1650)$, $N^\star(1710)$, $N^\star(1880)$, and $N^\star(1895)$ and their decays into $N\eta$ mode are taken into account, which are calculated using phenomenological Lagrangians~\cite{SajjadAthar:2022pjt}.
In the resonance sector, the various parameters appearing in the vector current sector  are fixed by applying this model to study the photon and  electron induced eta production from the free nucleon~\cite{Fatima:2023fez, Fatima:2022nfn}. We have fitted the coupling strength at the strong $R \rightarrow N\eta$ vertex and the electromagnetic coupling strength of the $N-R$ transition using the eta photoproduction data on the total cross section available from the MAMI collaboration~\cite{CrystalBallatMAMI:2010slt, A2:2014pie}. The $Q^2$ dependence of the electromagnetic $N-R$ transition form factors has been obtained by fitting the data of the electron induced $\eta$ production off the proton target for the total cross section at different values of $Q^2$~($Q^2 <1.4$~GeV$^{2}$) available from the CLAS collaboration~\cite{Denizli:2007tq}.
In the axial-vector sector, the axial-vector couplings have been calculated using the partially conserved axial-vector current~(PCAC) hypothesis and the generalized Goldberger-Treiman~(GT) relation with inputs from the experimentally determined strong $R \rightarrow N\pi$ couplings. 

In this work, we use a Fermi gas model in the local density approximation~(LDA) to describe the nuclear medium effects, which has been used earlier to study these effects in the electromagnetic sector~\cite{Chiang:1990ft, Carrasco:1993sk}. In this model, the differential scattering cross section  is given by~\cite{Hernandez:2013jka}:
   \begin{eqnarray}\label{eq:nucleus}
   \left( \frac{d\sigma}{dWdQ^2}\right)_{\nu A}&=&\int
 d^3r\sum_{N=n,p}2\int \frac{d^3p_N (r)}{(2\pi)^3}\ \Theta(E_F^N(r)-\nonumber \\
 && E_N)
 \,\Theta(E_N+q^0-E_\eta-E_F^{N'}(r)) \nonumber \\
 &&\left(\frac{d\sigma}{dWdQ^2}\right)_{\nu N},
   \end{eqnarray}
   where $\left(\frac{d\sigma}{dWdQ^2}\right)_{\nu N}$ is given in Eq.~(\ref{sigma:weak}), the factor of 2 is because of the spin degrees of freedom, $r$ is the radius of the nucleus, and $p_N$, $E_N$, respectively, are the momentum and energy of the initial nucleon.
The $\Theta$-function ensures that the momentum of the initial~(final) nucleon should be less~(more) than the Fermi momentum of the initial~(final) nucleon, which, in LDA, is a function of interaction point $r$ in the nucleus.

After integrating over the relevant variables, and with further kinematical simplifications,  we get
\begin{eqnarray}
 \frac{d\sigma}{dWdQ^2}
 &\approx& 2\pi \int r^2 dr \int d E_{\eta} d\cos \theta_\eta
\sum_{N=n,p}\hspace{-.15cm}\frac{G_F^2 a^2}{(2\pi)^6} |p_{\eta}|  \nonumber \\
&& \frac{\pi W}{32 E_{\nu}^2 E_N |\vec{P}|}(E_F^N(r)- E_{N}^{min}) \Theta(E_F^N(r)-E_{N}^{min})
\,\nonumber \\
&&\Theta(-P^2)\,\Theta(P^0)\, {\cal
  L}_{\mu\sigma}(k,k') {\cal
  W}^{\mu\sigma}(\tilde{p}_N,q,p_\eta),
\label{eq:gen1}
\end{eqnarray}
where ${P} = {q} - {p}_{\eta}$, $\tilde{p}_N = \sqrt{\tilde{E}_N^2 - M^2}$ is evaluated at $\tilde{E}_N = \frac{E_F^N(r) + E_N^{min}}{2}$, with
\begin{eqnarray}
 E_N^{min}&=&\max\{M,E_F^{N'} (r)-P^0,{\cal E}\}, \text{ and } \nonumber \\
 {\cal E}&=&\frac{-P^0+|\vec P|\sqrt{1-4M^2/P^2}}2.
\nonumber
\end{eqnarray}
${\cal   L}_{\mu\sigma} = \sum l_\mu l_\sigma^\dagger$, with $l_{\mu}$ given in Eq.~(\ref{lepton:current}), and ${\cal   W}^{\mu\sigma} = \sum J^\mu {J^\sigma}^\dagger$ is the hadronic tensor, given in Refs.~\cite{Fatima:2022nfn, Fatima:2023fez}.
The total scattering cross section $\sigma(E_{\nu_\mu})$ is obtained by integrating over the kinematical variable in Eq.~(\ref{eq:gen1}), where Eq.~(\ref{eq:gen1}) includes the effect of Fermi motion of the initial nucleon
and Pauli blocking of the final
nucleon.

The in-medium modification of the $N^\star(1535)$ resonance is evaluated using effective Lagrangians for the $\eta N$ interactions and many-body theory of nuclei, incorporating Fermi-sea effects and additional nuclear decay channels of $N^\star(1535)$ involving two nucleons. A well-established treatment of NME in this framework was provided by Chiang et al.~\cite{Chiang:1990ft}, where the self-energy of $N^\star(1535)$ in the nuclear medium was calculated using the empirical information about the nuclear dynamics.
The evaluation of the $\eta$-nucleus optical potential requires a reliable determination of the $N^\star(1535)$ self-energy $\Sigma$ in the nuclear medium. The real and imaginary parts of $\Sigma$ are known to be related with the mass and decay width of $N^\star(1535)$, respectively, in the nuclear medium~\cite{Chiang:1990ft}.
While the modification in the mass of $N^\star(1535)$ due to the self-energy correction leads to a minimal effect~\cite{Lehr:2003ka, Lehr:2003km}, the width undergoes considerable modification.
The
full decay width is replaced by the
density- and energy-dependent effective width following Ref.~\cite{Chiang:1990ft}:
\begin{equation}
    \Gamma_{N^\star}^{\rm free}(W)
    \longrightarrow
   \Gamma_{N^\star}^{\rm free}(W)+  \Gamma_{\rm eff}(\rho(r))
   \end{equation}
   where
   \begin{equation}
  \Gamma_{\rm eff}(\rho(r)) =
    \Gamma_{\eta}(\rho(r))
    +
    \Gamma_{\pi}(\rho(r))
    +
    \Gamma_{\pi\pi}(\rho(r)).
     \label{eq:gamma_eff}
\end{equation}
The individual terms in Eq.~(\ref{eq:gamma_eff}) for $\Gamma_{\rm eff}$ correspond to the decay width of
$N^\star(1535)\rightarrow N\eta$,
$N\pi$, and
$N\pi\pi$ decay channels, respectively, and the detailed expressions are given in Appendix.
The density dependence of the decay width incorporates the modification of the
available phase space and the additional collision-induced decay
and absorption processes generated by the surrounding nuclear
medium~\cite{Carrasco:1993sk}.

The propagation of the produced $\eta$ meson through the nucleus is treated semiclassically, with successive collisions explicitly taken into account.
An $\eta$ meson produced within a nucleus propagates through nuclear matter, where it can undergo elastic scattering or absorption via interactions with nucleons. To simulate this environment, nucleon positions are generated randomly from the center of the nucleus (defined as the origin). A minimum separation distance of 0.4~fm is maintained between any two nucleons to exclude short-range correlations. An $\eta$ meson is initialized at a random position \((x_0, y_0, z_0) \) inside the nucleus, propagating along the z-axis (0, 0, 1). The interaction rate is determined by the local nuclear density $\rho(r)$ and the channel-specific cross section $\sigma_{el}$ and $\sigma_{abs}$, for elastic scattering and absorption, respectively. The elastic cross section is calculated using partial-wave T-matrix elements by expanding the scattering amplitude in Legendre polynomials and relating the T-matrix to the respective phase shifts. The total scattering cross section $\sigma_T$ is calculated using the optical theorem, and $\sigma_{abs}$ is obtained by taking the difference between $\sigma_T$ and $\sigma_{el}$, which satisfactorily describes the experimentally data on the $\eta$ absorption cross section~\cite{Robig-Landau:1996prv}.
To determine the new direction \(\vec{n}^{\prime }\) of an elastically scattered \(\eta \) meson, we project its deflected trajectory onto a local coordinate system. First, we establish two mutually orthogonal unit vectors, \(\vec{v}\) and \(\vec{w}\), that are perpendicular to the initial direction vector \(\vec{n}\), then,
using the randomly sampled polar angle \(\theta \) and azimuthal angle \(\phi \), the new direction vector is calculated as: \(\vec{n^{\prime }}=\cos \theta \vec{n}+\sin \theta \cos \phi \vec{v}+\sin \theta \sin \phi \vec{w}\).

 To determine the \(\eta \) meson's fate (absorption or survival), its distance~($L$) to the nuclear surface is calculated. The next interaction point is sampled via: \(ds=-\lambda \ln p\) where \(p\) is a uniform random number, and \(\lambda = \left(\rho(r)\sigma_T\right)^{-1}\) is the mean free path.  If \(ds > L\) ,the \(\eta \) meson escapes the nucleus. If \(ds \le L\), the position updates (\(z_0 \leftarrow z_0 + ds\)). The absorption probability for $\eta$ is \(P_{abs} = \sigma_{abs}/\sigma_T\). If a uniform random number \(u < P_{abs}\), then the $\eta$ meson is absorbed. Otherwise, it undergoes elastic scattering. Repeating the procedure for many $\eta$ mesons gives the probability of elastic scattering, $P_\mathrm{el} = \frac{N_\mathrm{el}}{N_\mathrm{tot}}$  and the absorption probability $P_\mathrm{abs} = \frac{N_\mathrm{abs}}{N_\mathrm{tot}}$ where $N_\mathrm{el}$ and $N_\mathrm{abs}$ are the numbers of escaped and absorbed mesons, respectively and $N_\mathrm{tot} = N_\mathrm{el} + N_\mathrm{abs}$.

\begin{figure}
\begin{center}
\includegraphics[width=4cm,height=4.4cm]{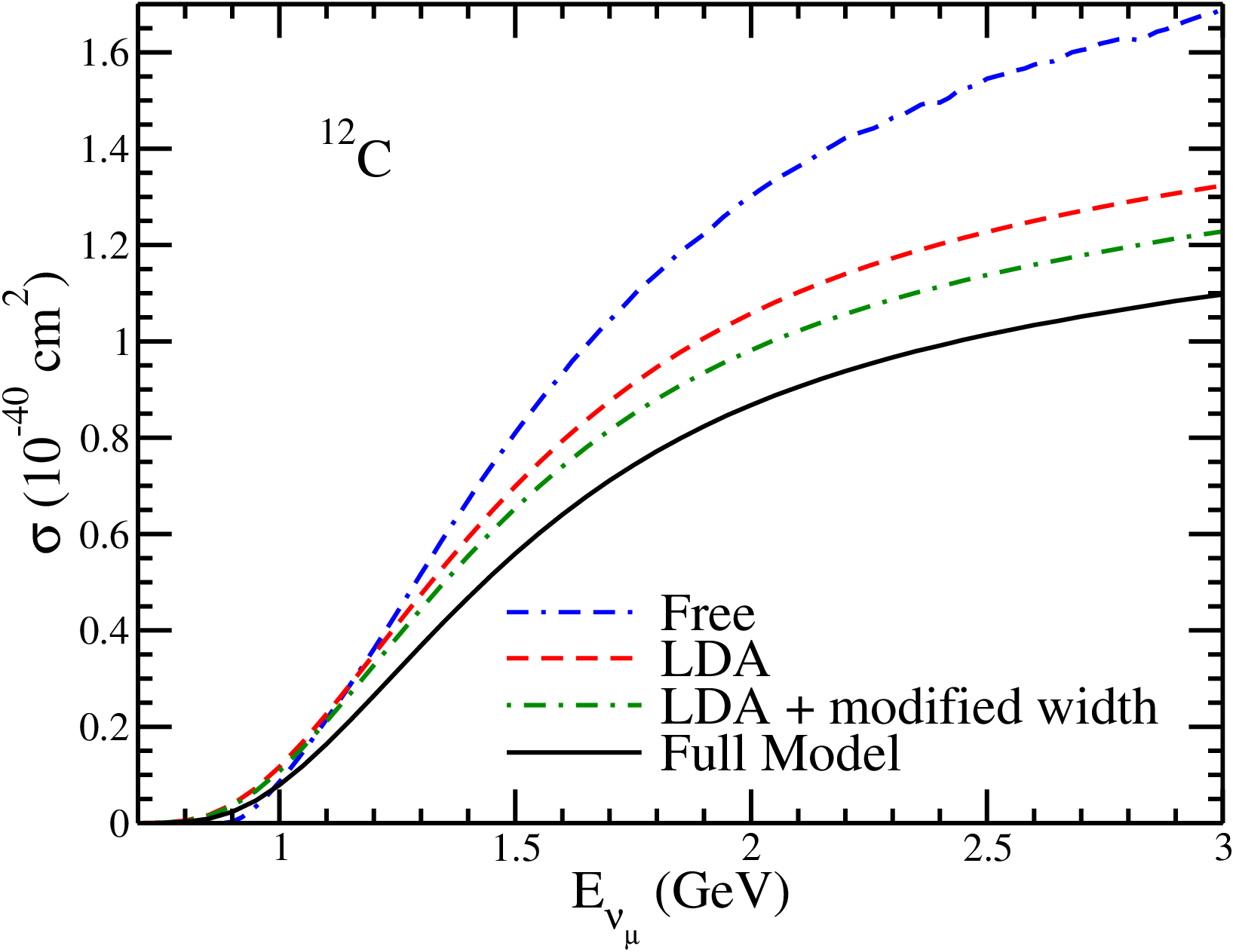}
\includegraphics[width=4cm,height=4.4cm]{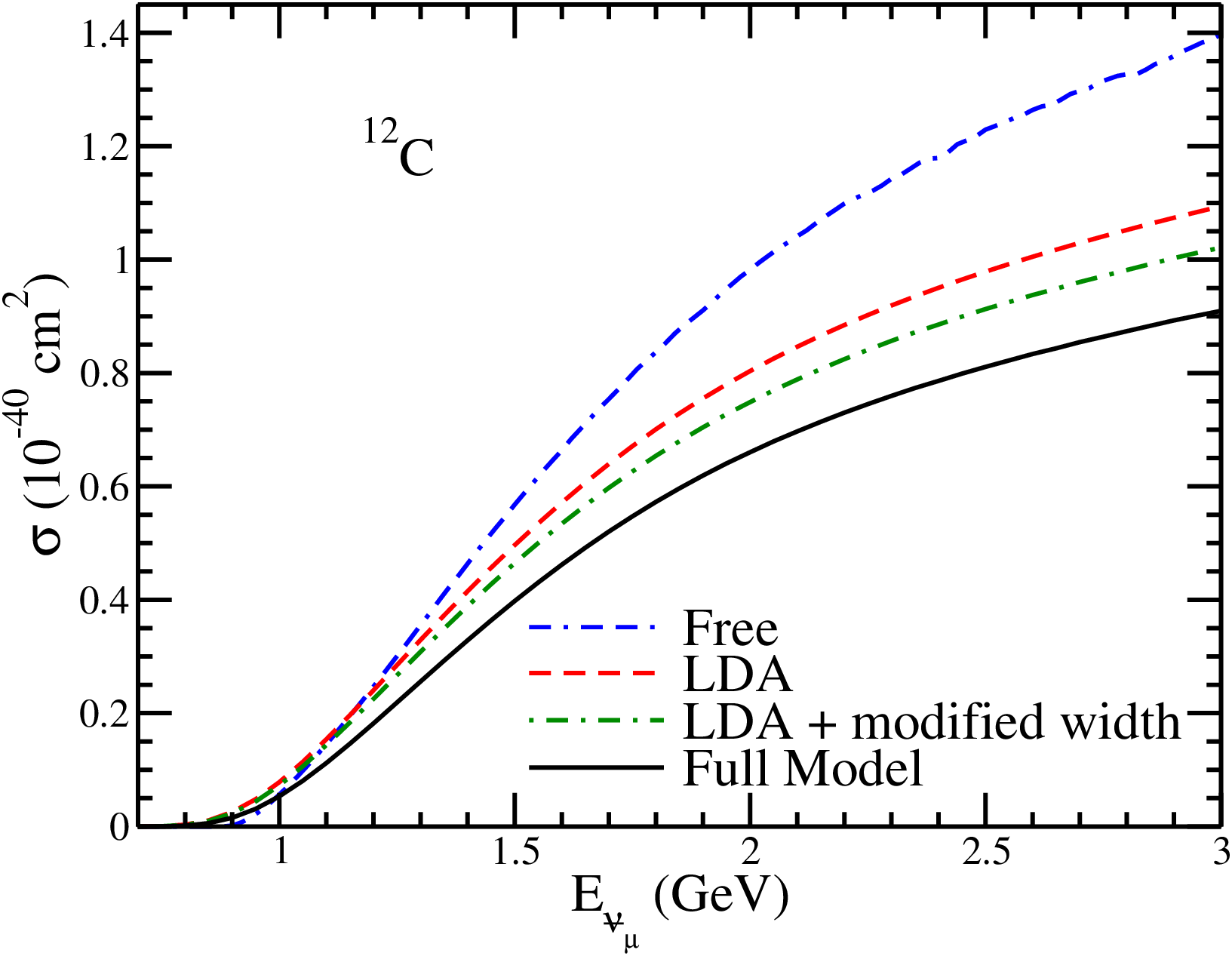}\\
\includegraphics[width=4cm,height=4.4cm]{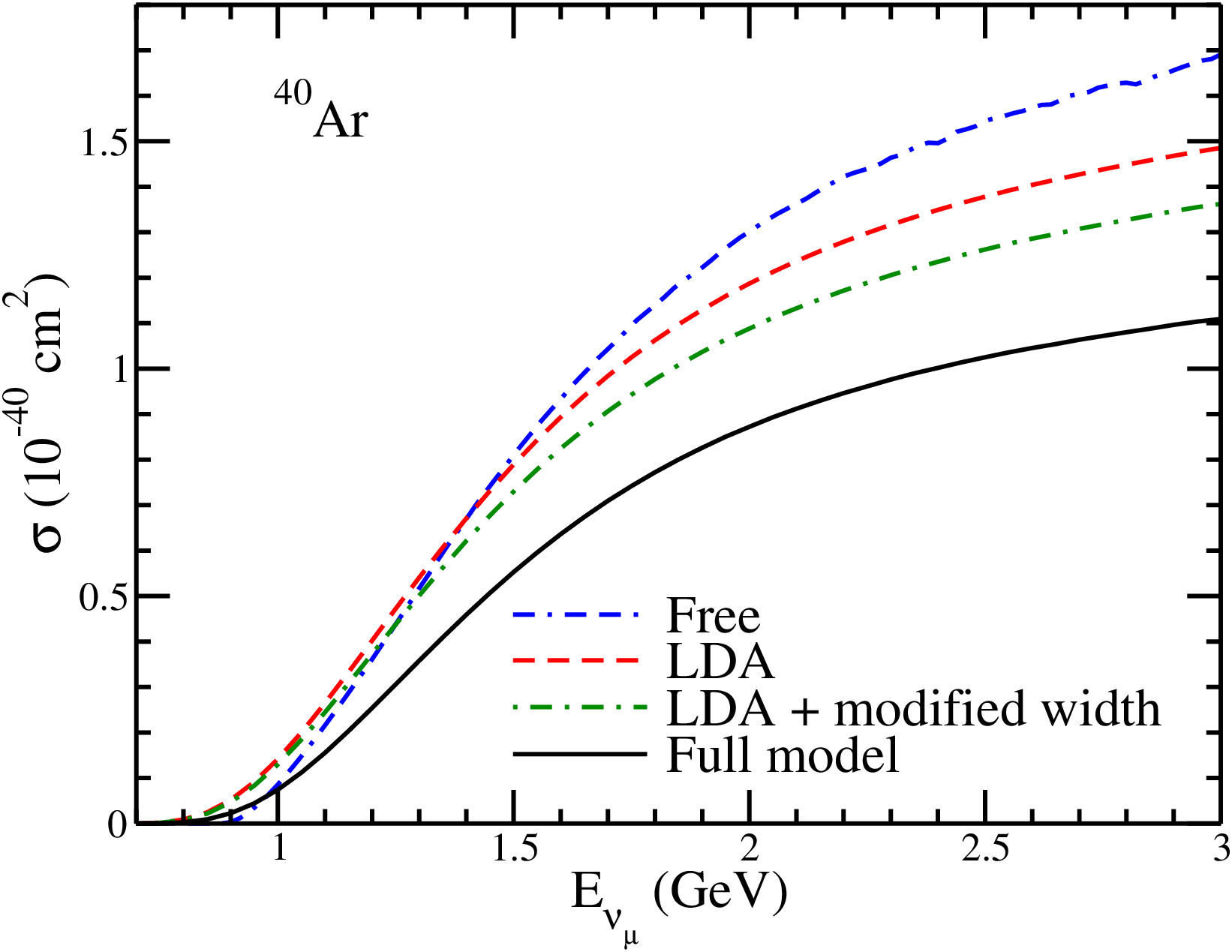}
\includegraphics[width=4cm,height=4.4cm]{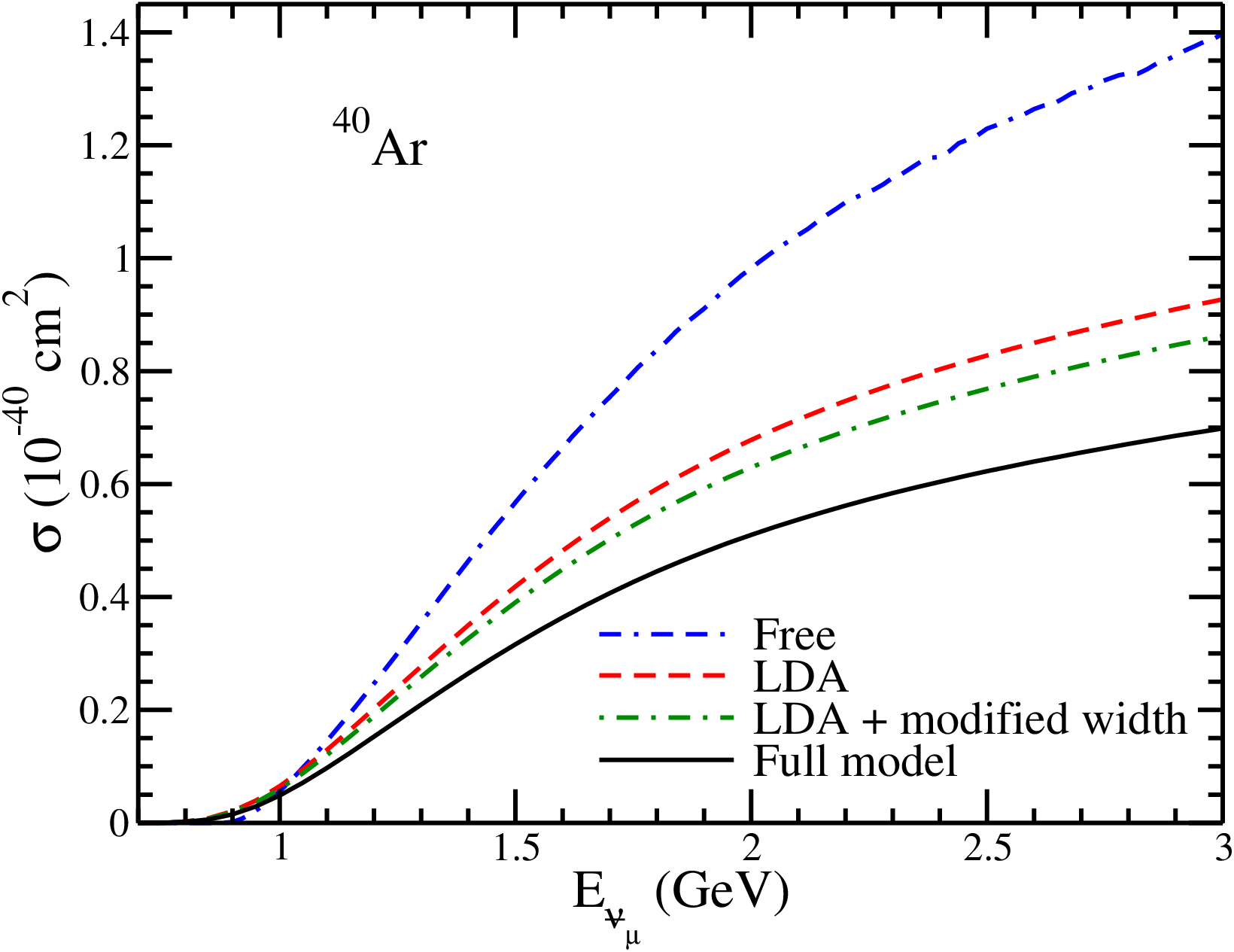}
\caption{$\sigma$ per interacting nucleon as a function of (anti)neutrino energy for the charged current neutrino~(left panel) and antineutrino~(right panel) induced $\eta$ production from $^{12}$C~(top panel) and $^{40}$Ar~(bottom panel) nuclear targets. Double-dash-dotted line shows the results for the free nucleon case. Dashed~(dash-dotted) line show the results of the NMEs evaluated in LDA~(LDA with $N^\star(1535)$ width modification). Solid line shows the results of the full model by including the effect of FSI and $\eta$ absorption.}
\label{fg_eta:sigma_carbon}
\end{center}
\end{figure}

\begin{figure}
\begin{center}
\includegraphics[width=4cm,height=4.4cm]{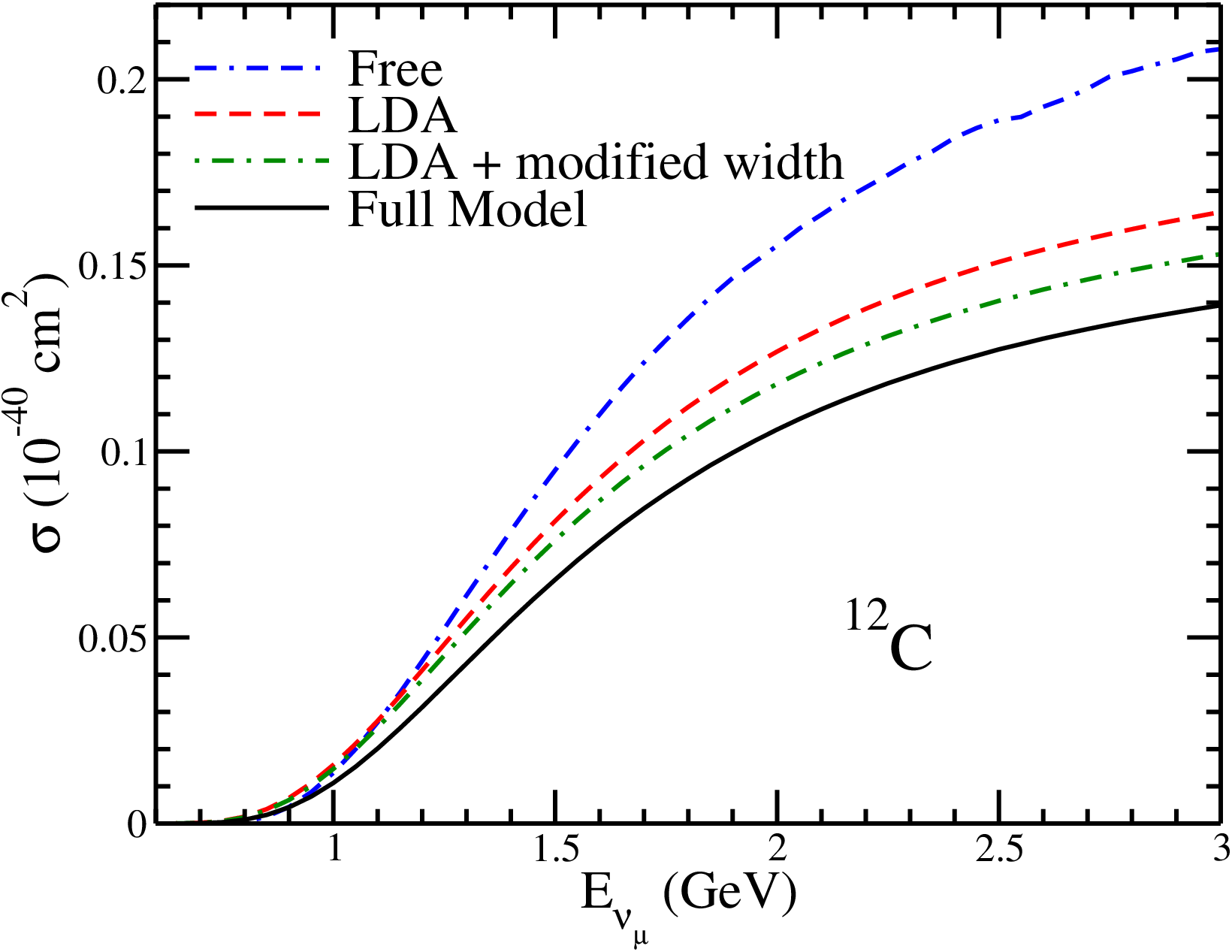}
\includegraphics[width=4cm,height=4.4cm]{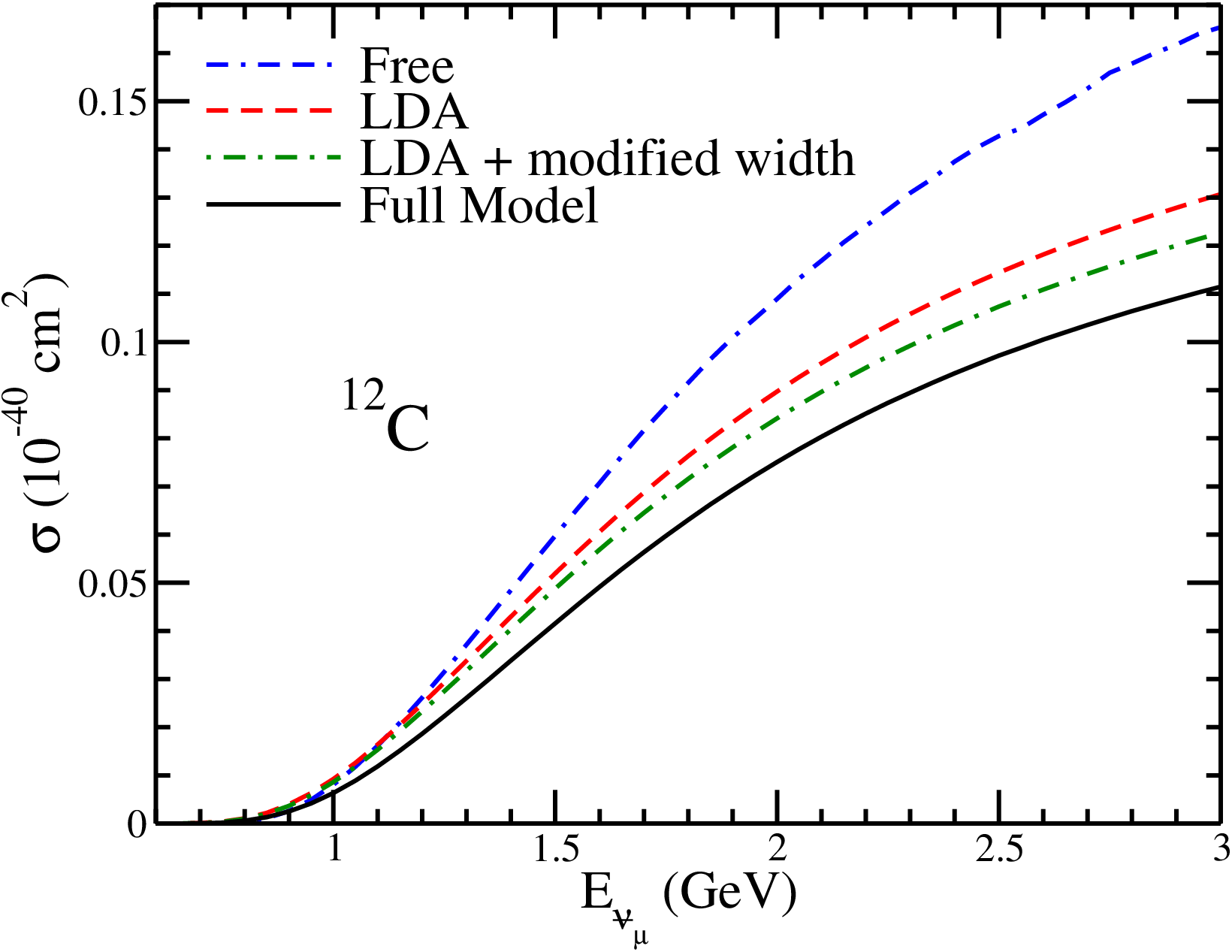}\\
\includegraphics[width=4cm,height=4.4cm]{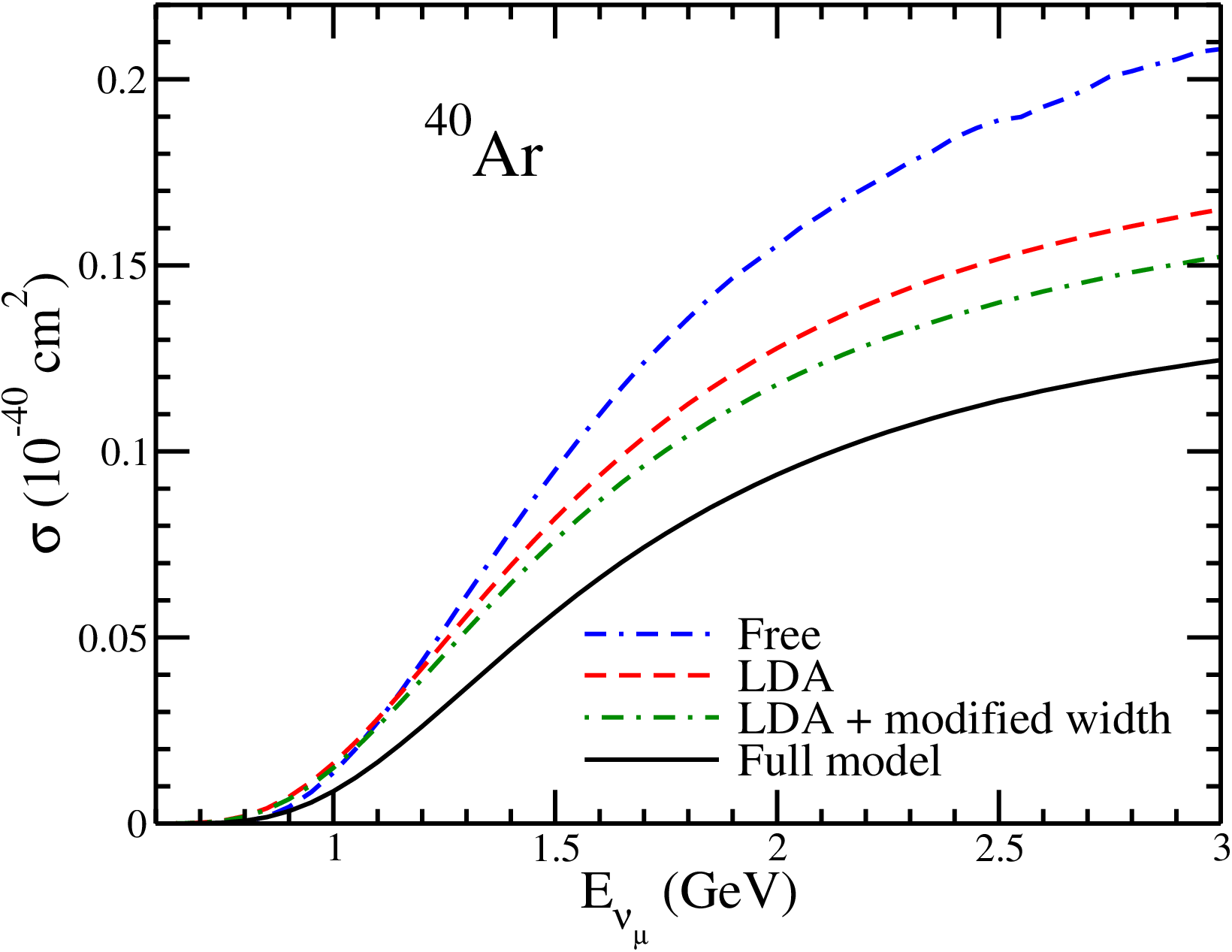}
\includegraphics[width=4cm,height=4.4cm]{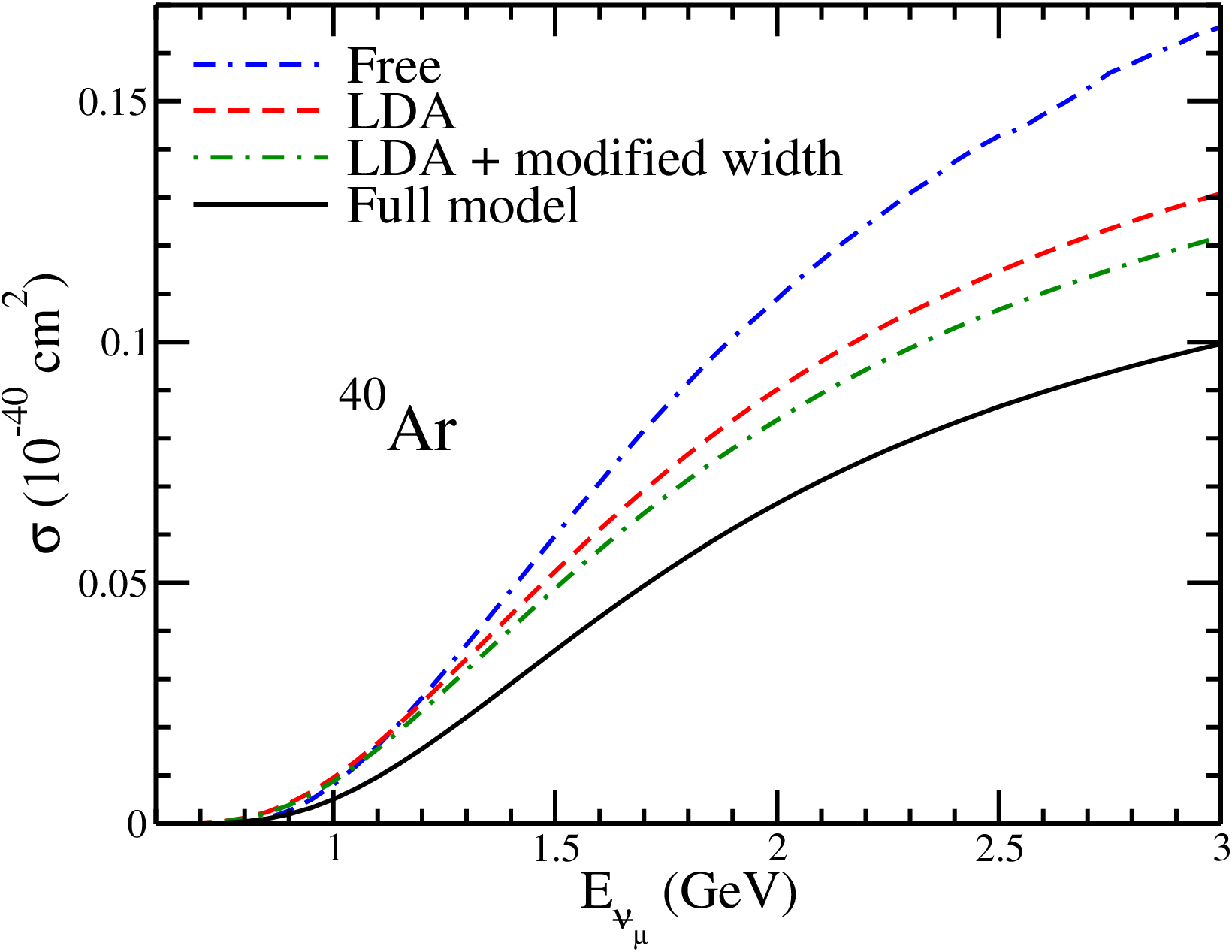}
\caption{$\sigma$ per interacting isoscalar nucleon as a function of (anti)neutrino energy for the neutral current neutrino~(left panel) and antineutrino~(right panel) induced $\eta$ production from $^{12}$C~(top panel)  and $^{40}$Ar~(bottom panel) nuclear targets. Lines and points have the same meaning as in Fig.~\ref{fg_eta:sigma_carbon}.}
\label{fg_eta:sigma_carbon1}
\end{center}
\end{figure}

\textit{ Results and Discussion - }
 In Figs.~\ref{fg_eta:sigma_carbon} and \ref{fg_eta:sigma_carbon1}, we present the results for the CC and NC induced $\nu_\mu(\bar\nu_\mu)$ interaction cross sections $\sigma(E_\nu)$ per interacting nucleon on $^{12}C$ and $^{40}Ar$ nuclear targets as a function of (anti)neutrino energies by integrating Eq.~(\ref{eq:gen1}) over the kinematical variables, respectively.  These results are obtained for the (anti)neutrino interactions (i)~off free nucleon target shown by the double-dash-dotted line, (ii)~off the nuclear target using~(a)~LDA where Fermi motion and Pauli blocking effects are incorporated shown by the dashed line, (b)~including medium modifications on the decay width of $N^\star(1535)$ resonance shown by the dash-dotted line, and (c)~also incorporating the multiple scattering and the FSI effects, shown by the solid line, which is the result of our full model.
It may be noted that the Fermi motion allows $\eta$-mesons to be produced even below the free nucleon production threshold.

In the case of CC neutrino induced $\eta$ production~(Fig.~\ref{fg_eta:sigma_carbon}), the cross section $\sigma_{CC}$ per interacting nucleon, obtained using the LDA with modified width increases by about 25\% (50\%) for $^{12}${C} ($^{40}${Ar}) at $E_{\nu_\mu}=1$~GeV as compared to the free nucleon case. With the increase in neutrino energy, however, the cross section decreases from the free case, and the cross over happens around 1.1~(1.25)~GeV in $^{12}$C~($^{40}$Ar). The reduction relative to the free nucleon results is $\approx$~25\%~(16\%) and 35\%~(20\%) for $^{12}$C~($^{40}$Ar) at $E_{\nu_{\mu}}=2$ and 3 GeV, respectively.
 For antineutrino induced reactions, the NME initially enhances the cross section by approximately 29\% (7\%) for $^{12}${C} ($^{40}${Ar}), followed by a cross over at 1.1~(1.03)~GeV, and
 a reduction of $\approx$~24\% (35\%) and 27\% (38\%), respectively, at $E_{\nu_{\mu}}=2$ and 3 GeV. When FSI effects and $\eta$-meson absorption are included, corresponding to our full model, the cross section is suppressed for all neutrino energies. The reduction in $\sigma_{CC}$ obtained using the full model relative to the results of the free case is about 8\%~(13\%) for $^{12}${C}~($^{40}$Ar) at $E_{\nu_\mu}=1$~GeV, which becomes 33\% and 35\% at 2 and 3~GeV, respectively, for both the nuclear targets.
Whereas, in the case antineutrino induced reactions, the reduction in $\sigma_{CC}$ becomes about 5\%~(35\%), 33\% (49\%), and 35\% (50\%) for $^{12}${C} ($^{40}$Ar)~at $E_{\nu_\mu}=1$, 2, and 3 GeV, respectively.

In the case of NC neutrino induced $\eta$ production~(Fig.~\ref{fg_eta:sigma_carbon1}), $\sigma_{NC}$ per interacting isoscalar nucleon, obtained using the LDA with modified width increases by about 7\% (10\%) for $^{12}${C} ($^{40}${Ar}) at $E_\nu=1$~GeV as compared to the free case. With the increase in neutrino energies, the cross section decreases, and the reduction relative to the free nucleon results is $\approx$~24\% and 27\%, respectively, for both $^{12}$C  and $^{40}$Ar at $E_{\nu_{\mu}}=2$ and 3 GeV. When FSI effects and $\eta$-meson absorption are included, corresponding to our full model, the cross section is suppressed for all energies, which is about 20\% (36\%), 32\% (39\%), and 33\% (40\%) for $^{12}${C} ($^{40}$Ar) at $E_{\nu}=1$, 2, and 3 GeV, respectively, as compared to the results obtained for the free nucleon case.
 In the case of NC antineutrino induced $\eta$ production, the results are  similar to what we observe in the case of NC neutrino induced $\eta$ production.

\begin{table}
\centering
\begin{tabular*}{80mm}{@{\extracolsep{\fill}}ccccc}\hline \hline
Experiment & \multicolumn{2}{c}{CC} & \multicolumn{2}{c}{NC} \\
& Nucleon & $^{40}$Ar & Nucleon & $^{40}$Ar \\ \hline
MicroBooNE & $1.87 $ & $1.30 $ & $0.23 $ & $0.15 $\\
SBND & 1.15 & 0.82 & 0.14  & 0.088 \\
DUNE & 12.1 & 8.14  & 1.49  & 0.9 \\
 \hline \hline
\end{tabular*}
\caption{$\langle \sigma \rangle$~(in units of $10^{-41}$~cm$^2$) in the nucleon case and full model, for the MicroBooNE, SBND, and DUNE experiments.}
\label{tab:param-p2}
\end{table}
We have obtained the flux averaged total scattering cross section $\langle \sigma \rangle$ by folding the cross section $\sigma(E_{\nu_\mu})$ over the MicroBooNE~\cite{MicroBooNE:2019nio}, SBND~\cite{Linyan:2026}, and DUNE~\cite{DUNE:2020jqi} fluxes~($\Phi (E_{\nu_\mu})$). For this we define
 \begin{eqnarray}\label{sigma:flux}
  \langle \sigma \rangle &=& \frac{\int \sigma(E_{\nu_\mu}) \Phi (E_{\nu_\mu}) dE_{\nu_\mu} }{\int \Phi (E_{\nu_\mu}) dE_{\nu_\mu}}.
\end{eqnarray}
The MicroBooNE and SBND experiments at the Fermilab use the Booster Neutrino Beam~(BNB) $\nu_\mu$ flux peaked around 0.7~GeV.  The planned DUNE experiment also at the Fermilab has an expected  neutrino flux that spans a broad energy range, with  a peak around 2.5 GeV, and a long tail. This broad spectrum could necessitate an appropriate cut on  $W$ and $Q^2$ to delineate the kinematic region of deep inelastic scattering~(DIS) contributing to the cross section. In particular, the moderate-W region~($W\leq$ 2 GeV, and the region of $W >2$~GeV with $Q^2 \le 1$~GeV$^2$), could play an important role as they probe the transition between resonance production and DIS, as discussed
recently by Athar and Morfin~\cite{SajjadAthar:2020nvy}. However, it has been found that the effect of the aforementioned $W$ and $Q^2$ restrictions hardly change~($< 0.1 \%$) the flux averaged cross section than the results obtained without any kinematical constrain.

Table-\ref{tab:param-p2} presents the results of $\langle \sigma \rangle$, averaged over MicroBooNE, SBND and DUNE fluxes, for the $\nu_\mu$ CC~(NC) induced scattering on the free neutron~(nucleon) target, and in $^{40}Ar$, with NME and FSI corresponding to our full model. It may be observed that the total reduction in $\langle \sigma \rangle$ due to NME and FSI  is $\sim 30\%$ at MicroBooNE and SBND energies, while this reduction is $\sim 33\%$ at DUNE energies. Comparing the theoretical results for the total~(CC+NC) $\eta$-production obtained in the full model, i.e., $\langle \sigma \rangle=1.45 \times 10^{-41}$~cm$^2$/nucleon,
with the reported results by the MicroBooNE collaboration~\cite{MicroBooNE:2023ubu}, which is $3.22 \pm 0.88 \pm 0.86 \times 10^{-41}$~cm$^2/$nucleon, we find that the theoretical calculations underestimate the experimental results. Thus, to conclude more study is required both theoretically as well as experimentally to understand the MicroBooNE data.

\textit{ Summary and conclusions - }
The results highlight the combined effects of Fermi motion, Pauli blocking, renormalization of the $N^\star(1535)$ properties, and FSI, leading to an overall suppression of $\sigma$. This  suppression is less pronounced at low (anti)neutrino energies, especially in CC interactions, underscoring the distinct impact of nuclear medium effects on CC and NC reactions. The magnitude of this suppression demonstrates that nuclear effects are not a minor correction but are essential for a quantitative description of neutrino and antineutrino induced $\eta$ production in nuclei. These results provide timely theoretical input for the interpretation of MicroBooNE and SBND measurements and, importantly, establish a necessary framework for controlling nuclear-model uncertainties in DUNE~\cite{DUNE:2020jqi} and other next-generation high-precision neutrino experiments. Furthermore,  weak $\eta$-production induced by atmospheric neutrinos constitutes an important physics input for the background estimation in proton decay searches, motivating improved theoretical predictions of the corresponding cross sections in the few GeV energy region~\cite{Super-Kamiokande:2024qbv}.

\textit{ Acknowledgements - }
AF acknowledges Department of Science and Technology~(DST), Government of India, for financial support under WISE Post-Doctoral Fellowship~(File no.~DST/WISE-PDF/PM-1/2025), to carry out this work, and Department of Physics, Aligarh Muslim University, Aligarh for providing the necessary facilities to pursue the research work.
 MSA is thankful to the DST for providing financial assistance under Grant No.
SR/MF/PS-01/2016-AMU.

\textit{ Data Availability - }
The data are not publicly available. The data are
available upon reasonable request from the authors.

\textit{ Appendix - }
The expressions of the partial decay widths $\Gamma_{\rm eff}(\rho(r))$ correspond to the decay of
$N^\star(1535)$ resonance into $N\eta$,
$N\pi$, and
$N\pi\pi$ decay channels are given by~\cite{Chiang:1990ft}:
 \begin{eqnarray}
  \frac{\Gamma_{\eta}}{2} &=& \frac{g_{\eta}^2}{4\pi} \frac{f_{\eta}^{2}}{m_\pi^2} \rho M \tilde{q}^3 \left[\frac{1}{q_0^2 - \tilde{q}^2 - m_{\eta}^2} \right]^2 F_{\eta}^4 (q_0, \tilde{q}), \\
  \frac{\Gamma_{\pi}}{2} &=& 3 \frac{g_{\pi}^2}{4\pi} \frac{f_{\pi}^{2}}{m_\pi^2} \rho M \tilde{q}^3 \left[\frac{1}{q_0^2 - \tilde{q}^2 - m_{\pi}^2} \right]^2 F_{\pi}^4 (q_0, \tilde{q}),\\
  \frac{\Gamma_{2\pi}}{2} &=& 6 \frac{1}{(2\pi)^3}\frac{f_{\eta}^{2}}{m_\pi^2} \left[\frac{C}{4m_{\text{red}}} \right]^2 \rho \int_{0}^{q_2^{\text{max}}} q_2 dq_2 \nonumber \\
  && \int_{0}^{q_1^{\text{max}}} q_1 dq_1 \left[2Mp_\eta^0 - q_1^2 - p_\eta^2 -2M\omega(q_2) \right]\nonumber \\
  && \frac{M}{\omega(q_2)} q_1^2 F_{\pi}^4 (q_1^0, q_1) \left[\frac{1}{q_1^0 - q_1^2 - m_{\pi}^2} \right]^2 \nonumber \\
  && P_{2} (q_1,q_2),
 \end{eqnarray}
 where $g_{\eta} = 1.613$, $f_{\eta} = 0.59$, $g_{\pi} = 0.564$, and $f_{\pi}=1$.
 The momentum $q_0$ and $\tilde{q}$ is given by
 \begin{equation}
  \tilde{q} = \sqrt{Mp_\eta^0 - |\vec{p}_\eta|^2/2}, \qquad q_0 = \frac{\tilde{q}^2}{2M}.
 \end{equation}
 The strong form factor $F_{i} (q_0, \tilde{q})$ is expressed as:
\begin{equation}
 F_{i} (q_0, \tilde{q}) = \left[\frac{\Lambda_i^2 - m_i^2}{\Lambda_i^2 - q_0^2 + \tilde{q}^2} \right], i=\pi,\eta
\end{equation}
with $\Lambda_{\pi} = 1.3$~GeV and $\Lambda_\eta =1.5$~GeV.

 \bibliographystyle{apsrev}

\end{document}